\documentclass[epjc3]{svjour3}
\RequirePackage[T1]{fontenc}

\usepackage{epsfig}
\usepackage{graphicx}

\newcommand{\Pomeron}{I\!\!P}
\def\p{I\!\!P}

\journalname{Eur. Phys. J. C}

\begin{document}


\title{A fresh look at factorization breaking in diffractive photoproduction of dijets at HERA
at next-to-leading order QCD}

\author{V. Guzey\thanksref{e1,addr1}
\and
M. Klasen\thanksref{e2,addr2}
}
\thankstext{e1}{e-mail: vguzey@pnpi.spb.ru}
\thankstext{e2}{e-mail: michael.klasen@uni-muenster.de}
\thankstext{c}{MS-TP-16-15}
\institute{National Research Center ``Kurchatov Institute'',
Petersburg Nuclear Physics Institute (PNPI), Gatchina, 188300, Russia\label{addr1} \and Institut f\"ur Theoretische Physik, Westf\"alische Wilhelms-Universit\"at M\"unster,
Wilhelm-Klemm-Stra{\ss}e 9, D-48149 M\"unster, Germany\label{addr2}
}

\date{\today}

\maketitle

\begin{abstract}

We calculate the cross section of diffractive dijet photoproduction in $ep$ scattering at
next-to-leading order (NLO) of perturbative QCD (pQCD), which we supplement by a model of
factorization breaking for the resolved-photon contribution. In this model, the suppression
depends on the flavor and momentum fraction of the partons in the photon. We show that
within experimental and theoretical uncertainties, the resulting approach provides a good
description of the available HERA data in most of the bins. Hence, taken together with the
observation that NLO pQCD explains well the data on diffractive photoproduction of open
charm in $ep$ scattering, our model of factorization breaking presents a viable alternative
to the scheme based on the global suppression factor.

\end{abstract}

\section{Introduction}
\label{sec:intro}

One of the highlights of the physics results obtained at HERA was the measurement of
inclusive diffraction $\gamma^{\ast} p \to X p$ in lepton-proton deep-inelastic scattering
(DIS). Combined analyses of the H1 and ZEUS experiments have been published in Ref.\
\cite{Aaron:2012hua}. Contrary to the expectations that the probability of large rapidity
gap events in DIS should be very small~\cite{Levy:1997pz}, it was found that diffractive
events in DIS constitute approximately $10$-$15$\% of the total cross section over a wide
range of $Q^2$. Furthermore, the QCD collinear factorization theorem for hard inclusive
diffraction~\cite{Collins:1997sr} allowed one to treat diffractive DIS on the same footing
as inclusive DIS. One can thus first introduce universal diffractive parton distribution
functions (dPDFs) and then determine them by fitting to the measured diffractive structure
function~\cite{Chekanov:2004hy,Aktas:2006hy,Aktas:2006hx}. The universality of the resulting
dPDFs is confirmed by a good agreement between the perturbative QCD (pQCD) calculations and
the data on diffractive production of dijets~\cite{Aktas:2007hn} and open
charm~\cite{Aktas:2006up} in DIS.

At the same time, in diffractive photoproduction of dijets in $ep$ scattering, based on the
well-known factorization breaking in diffractive dijet production in $\bar{p}p$ collisions
at the Tevatron~\cite{Affolder:2000vb,Kaidalov:2001iz,Klasen:2009bi}, collinear factorization
is generally not expected to hold \cite{Kaidalov:2003xf,Klasen:2004qr,Klasen:2004ct}. However,
the pattern of this factorization breaking remains an open question \cite{Klasen:2005dq}; for
recent reviews see, e.g., Refs.\ \cite{Klasen:2008ah,Kaidalov:2009fp,Klasen:2010vk}. To
summarize, while the recent H1~\cite{Aaron:2010su,Cerny} and ZEUS~\cite{Chekanov:2007rh} data
on diffractive dijet photoproduction are largely consistent with each other after the
renormalization of the ZEUS data, the next-to-leading order (NLO) perturbative QCD calculations
overestimate the data by approximately $40$-$50$\%. The theory and the data can be made
consistent by introducing either a global suppression factor of $0.5$ or a suppression
factor of approximately $0.4$ only for the resolved-photon contribution. The most recent H1
measurement of diffractive dijet photoproduction with a leading proton~\cite{Andreev:2015cwa}
is also consistent with the observation that NLO pQCD globally overestimates the data by the
factor of $0.5$.

Factorization breaking in diffractive dijet photoproduction is a result of soft inelastic
photon interactions with the proton, which populate and thus partially destroy the final-state
rapidity gap. This effect is usually described in the literature by a rapidity gap survival
factor $S^2 \leq 1$. Since the magnitude of $S^2$ decreases with an increase of 
the interaction strength between the probe and the target,
the pattern of the factorization breaking can be related to various components of 
the photon~\cite{Kaidalov:2009fp}. In the laboratory reference frame, the high-energy photon
interacts with hadronic targets by fluctuating into various configurations (components)
interacting with the target with different cross sections. These fluctuations contain both
weakly-interacting (the so-called point-like) components and the components interacting with
large cross sections, which are of the order of the vector meson-proton cross sections.
This general space-time picture of photon-hadron interactions at high energies 
is usually realized in the framework of such approaches as the vector meson dominance (VMD) model
and its generalizations~\cite{Bauer:1977iq} or the color dipole model~\cite{Nemchik:1996pp,Kowalski:2006hc}. 
It is also used in the language of collinear factorization, where 
the photon structure function and parton distribution functions (PDFs) are given by a sum of the resolved-photon contribution
corresponding to the VMD part of the photon wave function and the point-like (inhomogeneous) term
originating from the $\gamma \to q \bar{q}$ splitting, see, e.g., Ref.~\cite{Gluck:1991jc}.
Note that the direct-photon contribution to diffractive dijet photoproduction corresponds to
the configurations interacting with very small cross sections of the order of $1/E_T^2$ ($E_T$ is the transverse jet
energy), which preserves factorization.

Let us recall 
that the hadron (VMD) part of the photon PDFs contributes only for small $x_{\gamma}$, whereas 
the point-like term gives the dominant contribution for large $x_{\gamma}$.
Here, $x_{\gamma}$ is the light-cone momentum fraction of a parton in the photon.
Based on the arguments presented above, it is then natural to expect $S^2=1$ for the direct-photon
contribution localized near $x_{\gamma}=1$, $S^2 \approx 0.34$ for the hadron-like component of
the photon at small $x_{\gamma}$, and $S^2 \approx 0.53-0.75$ for the gluon and quark contributions
at large $x_{\gamma}$ corresponding to small, but non-negligible factorization breaking due to the
point-like component of the resolved photon \cite{Kaidalov:2009fp,Klasen:2010vk}.
Note that in pQCD the separation of the direct and the resolved photon
contributions is unambiguous only at leading order. At NLO, it becomes a matter of convention
depending on the factorization scheme and the factorization
scale~\cite{Klasen:1994bj,Klasen:1995ab,Klasen:2002xb}.  

Another important observation relevant to the possible pattern of factorization breaking is
that the HERA data on diffractive photoproduction of open charm~\cite{Chekanov:2007pm} agree
with the NLO pQCD calculations~\cite{Frixione:1995qc,Klasen:2011ax}, and hence no factorization
breaking is required. This calls into question the assumption of a global suppression factor
modeling factorization breaking and indicates that $S^2$ for the resolved-photon contribution
may depend on the parton flavor. In particular, $S^2 \approx 1$ for the charm quark distribution
in the photon, which agrees with the observation that in the 
VMD model
the $J/\psi$-proton cross section is of the order of a few mbarn, i.e.\ much smaller than the
$\rho$-nucleon cross section~\cite{Bauer:1977iq}.

In the present work we revisit the issue of factorization breaking in diffractive dijet
photoproduction in $ep$ scattering and perform NLO pQCD calculations of the corresponding
cross sections, which we combine with a new flavor-dependent and momentum fraction-dependent
scheme of factorization breaking for the resolved-photon contribution. We demonstrate that the
results of our calculations provide a good description of the H1~\cite{Aaron:2010su,Cerny} and
ZEUS~\cite{Chekanov:2007rh} data on diffractive dijet photoproduction in $ep$ scattering at
HERA, while simultaneously, by construction, not conflicting with the good pQCD description of
diffractive photoproduction of open charm in $ep$ scattering. Thus, next-to-leading order
perturbative QCD coupled with the physically motivated assumption about the rapidity gap
survival probability for the resolved-photon contribution and the effect of hadronization
corrections provide a good description of all available HERA data on diffractive dijet
photoproduction. This result reinforces the conclusion of~Ref.~\cite{Kaidalov:2009fp}.

\section{New scenario for factorization breaking in diffractive photoproduction of dijets in $ep$}
\label{sec:fb}

\begin{figure}
 \centering
 \includegraphics[width=0.46\columnwidth]{fig1a}
 \includegraphics[width=0.49\columnwidth]{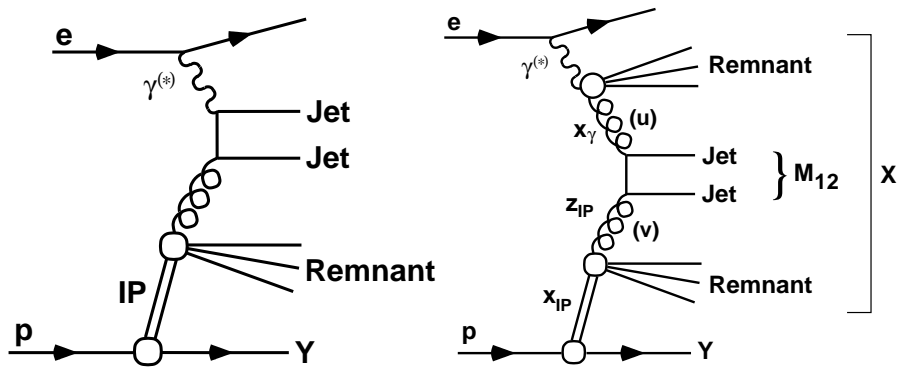}
 \caption{\label{fig:2}Diffractive production of dijets with invariant mass
 $M_{12}$ in direct (left) and resolved (right) photon-pomeron collisions,
 leading to the production of one or two additional remnant jets.}
\end{figure}

We performed next-to-leading order (NLO) pQCD calculations~\cite{Klasen:2010vk} of the cross
sections of diffractive photoproduction of dijets in $ep$ scattering 
$ep\to e+2{\rm jets}+ X^{\prime}+Y$
using the kinematic
conditions and cuts of the H1~\cite{Aaron:2010su,Cerny} and ZEUS~\cite{Chekanov:2007rh}
measurements of this process. 
This process is illustrated in Fig.\ \ref{fig:2}. As is well known,
%
%
jet production with real photons involves direct interactions of the
photon with quarks or gluons from the proton (or in our case from the pomeron)
as well as resolved photon contributions, leading to parton-parton
interactions and an additional remnant jet coming from the photon.
For the direct interactions, 
factorization is expected to be valid as in the case of DIS, whereas we expect 
it to fail for the resolved process as in hadron-hadron scattering. For this
part of photoproduction one would therefore naively expect a similar suppression factor
due to rescattering effects of the hadronic fluctuations of the photon.
The expression for the cross section reads: 
\begin{eqnarray}
d \sigma(ep\to e &+&2{\rm jets}+X^{\prime}+Y)
=\sum_{i,j} \int dt \int dx_{\Pomeron}
\int dz_{\Pomeron} \int dy \int dx_{\gamma}
\nonumber\\
&\times & S^2_i(x_{\gamma})f_{\gamma/e}(y)  f_{i/\gamma}(x_{\gamma},\mu^2) f^{D(4)}_{j/p}(x_{\Pomeron},z_{\Pomeron},t,\mu^2)
d \hat{\sigma}_{ij \to {\rm jets}}^{(n)} \,,
\label{eq:cs}
\end{eqnarray}
where $X'$ denotes the pomeron (and possibly photon) remnant jet(s);
$Y$ denotes either a proton or a low-mass proton excitation;
$S^2_i(x_{\gamma})$ is the factor modeling factorization breaking;
$f_{\gamma/e}(y)$ is the photon flux of the electron depending on the photon
 light-cone momentum fraction $y$;
$f_{i/\gamma}(x_{\gamma},\mu^2)$ is the PDF of parton $i$ in the photon
 depending on the momentum fraction $x_{\gamma}$ and  the factorization scale $\mu$;
$f^{D(4)}_{j/p}(x_{\Pomeron},z_{\Pomeron},t,\mu^2)$ is the diffractive PDF of the proton,
 which depends on $x_{\Pomeron}$ (the momentum fraction carried by the diffractive exchange
 or ``Pomeron'') and $z_{\Pomeron}$ (the momentum fraction of parton $j$ with respect to the
 ``Pomeron'' momentum), the invariant momentum transfer squared $t$, and $\mu^2$;
and $d \hat{\sigma}_{ab \to {\rm jets}}^{(n)}$ is the elementary pQCD cross section for
 the production of an $n$-parton final state in the interaction of partons $i$ and $j$.
The sum over $i$ involves both quarks and gluons (resolved-photon contribution) and the
 photon (direct-photon contribution).
For input in Eq.~(\ref{eq:cs}), we used the GRV photon PDFs transformed to the
$\overline{\rm MS}$ scheme~\cite{Gluck:1991jc} and the 2006 H1 proton diffractive PDFs
(fit B)~\cite{Aktas:2006hy}.

We 
explained in the Introduction that
the space-time picture of high-energy photon-proton interactions suggests that 
in diffractive dijet photoproduction on the proton, QCD diffractive factorization holds 
for the direct-photon contribution and is broken for the resolved-photon contribution. Moreover,
for the latter contribution, the factorization breaking is strongest at small $x_{\gamma}$, small, but
non-negligible, for large $x_{\gamma}$, and depends on the parton flavor. 
In the framework of collinear QCD factorization, the decrease of factorization-breaking effects
in the resolved-photon contribution with an increase of $x_{\gamma}$ can be explained by 
the observation that based on the factorization of the collinear singularity, there should be
a smooth transition from the resolved-photon contribution at large $x_{\gamma}$ to the direct-photon 
contribution.

Therefore, we model the effect of factorization breaking by introducing
the following suppression factor of $S_{i}^2(x_{\gamma})$ for the resolved-photon contribution (i.e.\
for the photon PDFs)
in Eq.~(\ref{eq:cs}):
\begin{equation}
S_{i}^2(x_{\gamma}) \to  \left\{\begin{array}{ll}
1 \,, & i=c \,, \\
A_q\, x_{\gamma} +0.34 \,, \quad & i=u,d,s \,,\\
A_g\, x_{\gamma}+0.34 \,, \quad &  i=g \,,\end{array} \right. 
\label{eq:S_j}
\end{equation}
where $i$ is the parton flavor; 
$A_q=0.37-0.41$ and $A_g=0.19-0.24$. The given ranges of values take into account the possible
effective dependence of $S_{i}^2(x_{\gamma})$ on the hard resolution scale, where the first and the
second values correspond to $E_T^{\rm jet 1}=5$ and 7.5 GeV, respectively. Thus, the factor of
$S_{i}^2(x_{\gamma})$ in Eq.~(\ref{eq:S_j}) represents a linear interpolation between the domain of
small $x_{\gamma}$ dominated by the hadronic contribution to photon PDFs, where $S_i^2(x_{\gamma})=0.34$, and
the regime of large-$x_{\gamma}$ dominated by the point-like contribution to photon PDFs, where
$S_q^2(x_{\gamma})=0.71-0.75$ for quarks and $S_g^2(x_{\gamma})=0.53-0.58$ for gluons, see Ref.\
\cite{Kaidalov:2009fp}.
Note that the model of Eq.~(\ref{eq:S_j}) assumes no factorization breaking in the charm quark channel according to
the observation that NLO pQCD describes well diffractive photoproduction of open charm in $ep$ scattering,
see the Introduction. 

%
%
\begin{table}
 \caption{Kinematic cuts applied in the most recent H1 \cite{Aaron:2010su,Cerny}
 and ZEUS \cite{Chekanov:2007rh} analyses of diffractive dijet photoproduction. \label{tab:1}}
\centering
 \begin{tabular}{c|c|c}
 H1 low-$E_T^{jet}$ cuts & H1 high-$E_T^{jet}$ cuts & ZEUS cuts\\
 \hline
 $Q^2<$ 0.01 GeV$^2$   &  $Q^2<$ 0.01 GeV$^2$ & $Q^2<$ 1 GeV$^2$  \\
 0.3 $< y <$ 0.65     &  0.3 $< y <$ 0.65 & 0.2 $< y <$ 0.85     \\
 $E_T^{\rm jet1}>$ 5 GeV & $E_T^{\rm jet1} >$ 7.5 GeV & $E_T^{\rm jet1} >$ 7.5 GeV \\
 $E_T^{\rm jet2}>$ 4 GeV & $E_T^{\rm jet2} >$ 6.5 GeV & $E_T^{\rm jet2} >$ 6.5 GeV \\
 $-1 < \eta^{\rm jet1(2)} < 2$ &  $-1.5 < \eta^{\rm jet1(2)} < 1.5$ & $-1.5 < \eta^{\rm jet1(2)} < 1.5$ \\
 $z_{\p} <$ 0.8    &  $z_{\p} <$ 1  \\
 $x_{\p} <$ 0.03      & $x_{\p} <$ 0.025 & $x_{\p} <$ 0.025 \\
 $|t| <$ 1 GeV$^2$    &  $|t| <$ 1 GeV$^2$ & $|t| <$ 5 GeV$^2$ \\
 $M_Y <$ 1.6 GeV      &  $M_Y <$ 1.6 GeV\\
 \end{tabular}
\end{table}
%

\begin{figure}[t]
\begin{center}
\epsfig{file=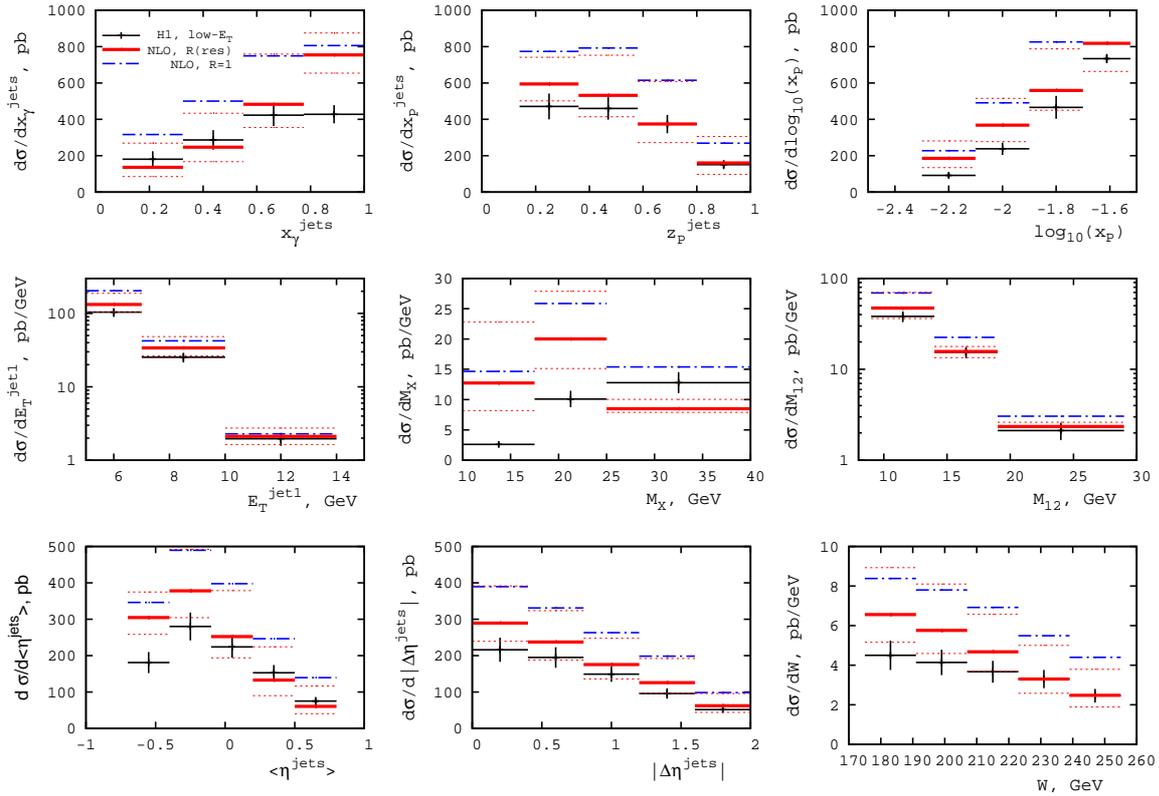,scale=0.8}
 \caption{Cross section of diffractive dijet photoproduction in $ep$ scattering: comparison of the NLO pQCD predictions combined with the model of factorization breaking of
 Eq.~(\ref{eq:S_j}) (red solid lines)
 to the H1 data with the low-$E_T^{\rm jet}$ cut~\cite{Aaron:2010su};
 the theoretical uncertainty due to the variation of the normalization and factorization scales is shown by the
 red dotted lines. Also, the NLO pQCD results without the effect of factorization breaking are given by the blue dot-dashed
 lines labeled ``NLO, $R=1$''. Note that the pQCD predictions include the hadronization corrections.
 }
 \label{fig:H1_2010_lj}
\end{center}
\end{figure}

\begin{figure}[t]
\begin{center}
\epsfig{file=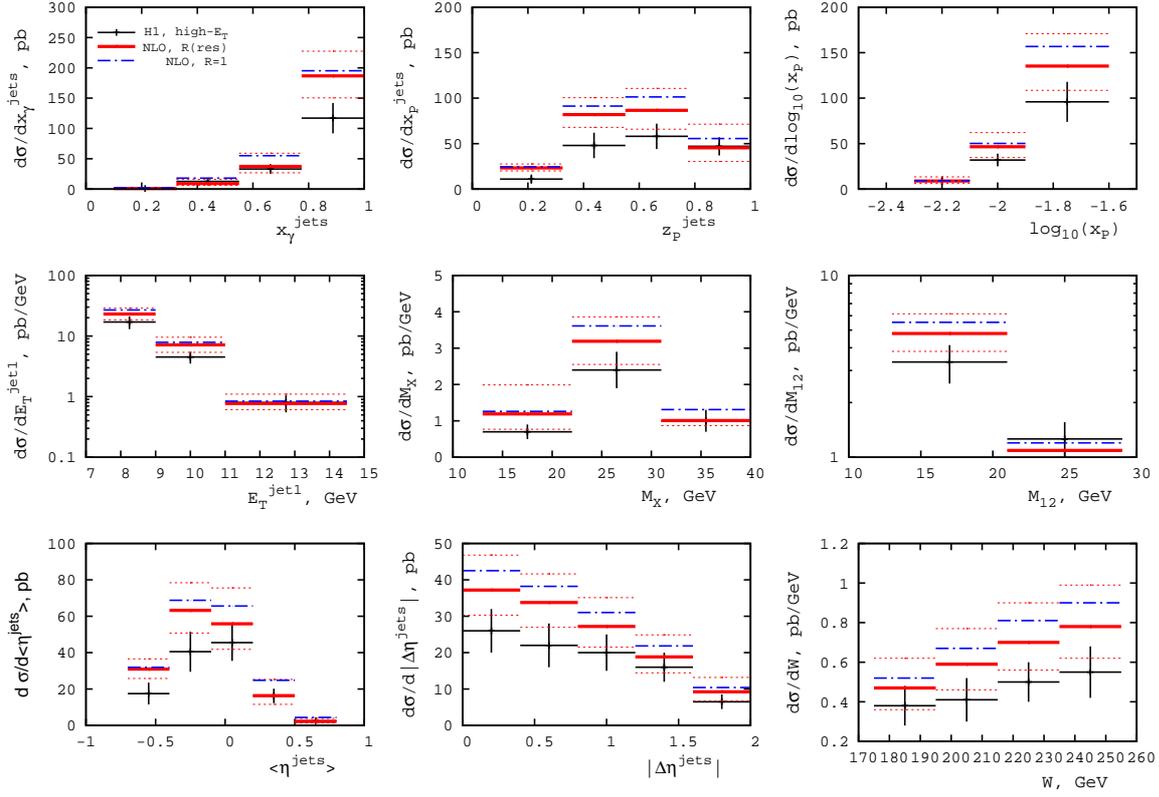,scale=0.8}
 \caption{The same as in Fig.~\ref{fig:H1_2010_lj}, but for the 
 H1 data with the high-$E_T^{\rm jet}$ cut~\cite{Cerny}.}
 \label{fig:H1_2010_hj}
\end{center}
\end{figure}

\begin{figure}[t]
\begin{center}
\epsfig{file=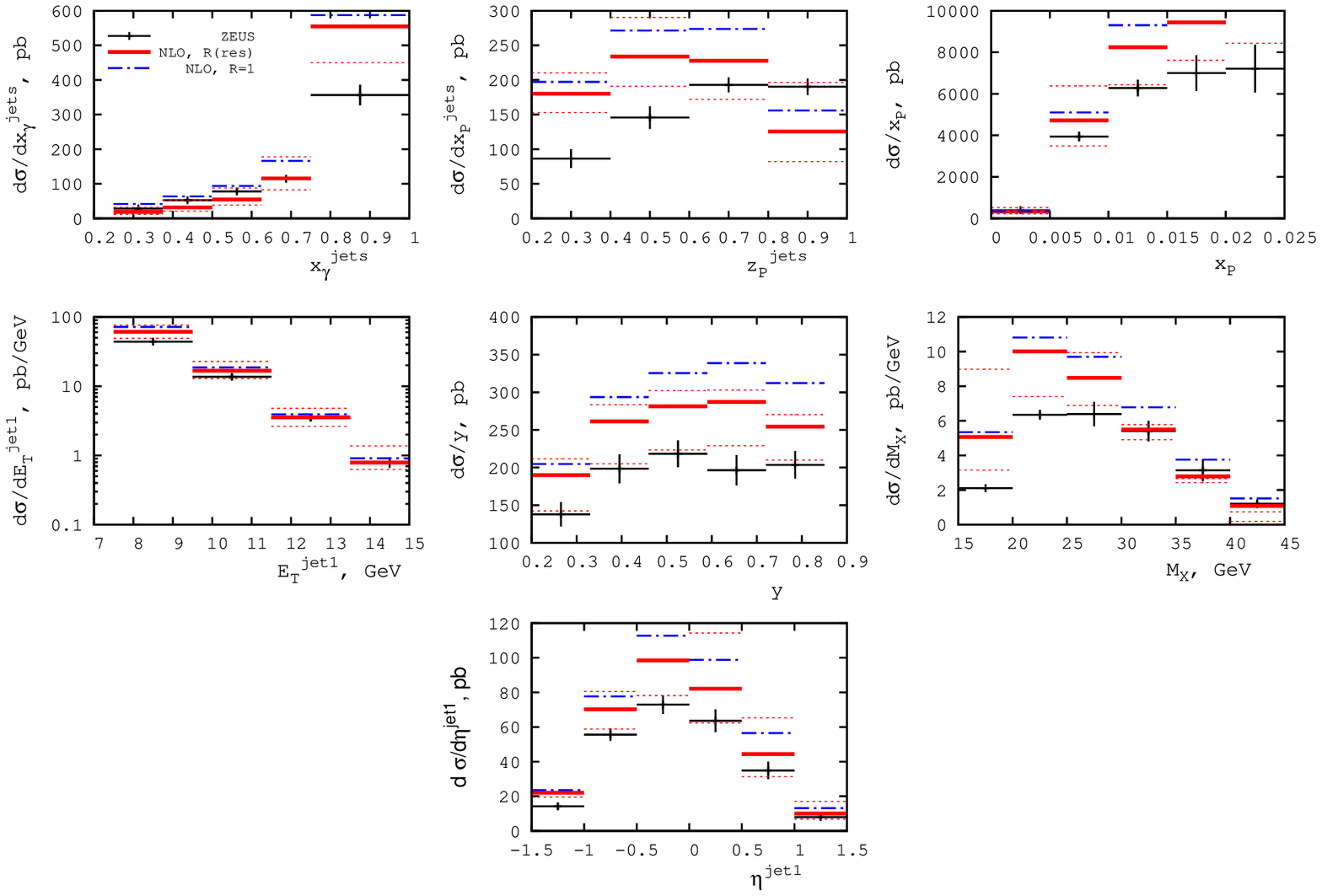,scale=0.8}
 \caption{The same as in Fig.~\ref{fig:H1_2010_lj}, but for the 
 ZEUS data~\cite{Chekanov:2007rh}.}
 \label{fig:ZEUS}
\end{center}
\end{figure}

The comparison of the results of our calculations to the H1 and ZEUS data is shown in
Figs.~\ref{fig:H1_2010_lj}, \ref{fig:H1_2010_hj} and \ref{fig:ZEUS}.
The kinematic cuts of the experimental analyses are summarized in Table~\ref{tab:1},
where $Q^2$ refers to the photon virtuality, $y$ its momentum fraction in the electron,
$E_T^{\rm jet1(2)}$ are the leading
and subleading transverse jet energies, and $\eta^{\rm jet1(2)}$ their rapidities.
In Figs.~\ref{fig:H1_2010_lj}--\ref{fig:ZEUS}, the thick red solid
lines correspond to the calculation, when the renormalization and factorization scale $\mu$
is identified with the transverse energy of jet 1, $\mu=E_T^{\rm jet 1}$. The thin red dotted
lines quantify the scale uncertainty of our NLO calculations and correspond to $\mu=2
E_T^{\rm jet 1}$ (lower) and $\mu=E_T^{\rm jet 1}/2$ (upper). For comparison, we also also show
the unsuppressed predictions (assuming no factorization breaking) by the blue dot-dashed
lines labeled ``NLO, $R=1$''. Note that our theoretical calculations have been
multiplied by the hadronization corrections in each bin~\cite{Aaron:2010su,Cerny,Chekanov:2007rh}. 
In the different panels, the values of the cross section are shown as functions of the following variables, see, e.g.,
Ref.~\cite{Aaron:2010su}:
$x_{\gamma}^{\rm jets} = \sum_{\rm jets} (E_i-P_{i,z})/(E_X-P_{X,z})$ is the hadron-level estimator 
of the parton momentum fraction in the photon, where the sum runs over the hadronic final states $i$ included in the jets
and $X$ refers to the full diffractive final state;
$z_{\Pomeron}^{\rm jets} = \sum_{\rm jets} (E_i+P_{i,z})/(E_X+P_{X,z})$ is the estimator of the ``Pomeron'' momentum
fraction carried by a parton;
$x_{\Pomeron}=(E_X+P_{X,z})/(2E_p^{\rm beam})$ is the measured ``Pomeron'' momentum fraction, where $E_p^{\rm beam}$ is the proton
beam energy; $E_T^{\rm jet 1}$ is the transverse energy of jet 1;
$M_X$ is the invariant mass of the diffractive final state (two jets plus diffractively-produced remnants of the photon 
and the ``Pomeron'');
$M_{12}$ is the invariant mass of the dijet system; $\langle \eta^{\rm jets} \rangle=(\eta_1+\eta_2)/2$ 
and  $\Delta \eta^{\rm jets}|=|\eta_1-\eta_2|$ are the average and the relative jet rapidities, and 
$W$ is the invariant photon-proton energy.

One can see from Figs.~\ref{fig:H1_2010_lj}-\ref{fig:ZEUS} that within theoretical
uncertainties, NLO perturbative QCD combined with the model of factorization breaking
of Eq.~(\ref{eq:S_j}) provides a good description of the data for most of the bins.
The quality of the data description is similar to that of Ref.~\cite{Klasen:2010vk},
where factorization breaking is realized either by the global or the resolved-only,
flavor- and $x$-independent suppression factors.

An inspection of Figs.~\ref{fig:H1_2010_lj}-\ref{fig:ZEUS} shows that NLO pQCD
correctly reproduces the shape of almost all considered distributions and only fails to
explain the normalization in some bins receiving a significant contribution from the
direct-photon contribution, which is unsuppressed in our factorization breaking scheme.
The most notable example is the distribution at high $x_{\gamma}^{\rm jets}$ at low $E_T^{\rm jet}$.
It was hypothesized in Ref.~\cite{Kaidalov:2009fp} that additional sizable hadronization 
corrections along with bin migration effects, which are not included in our analysis, 
might help to improve the agreement between theory and data at large $x_{\gamma}^{\rm jets}$.
This hypothesis is supported by very similar observations in inclusive photoproduction at
low $E_T^{\rm jet}$~\cite{Klasen:2002xb}.
 
Integrating our results for $d\sigma/dE_{T}^{\rm jet 1}$ over $E_{T}^{\rm jet 1}$, we obtain
the theoretical prediction for the integrated cross section $\sigma_{\rm NLO}^{\rm tot}$.
Table~\ref{table:sigma} gives our results for $\sigma_{\rm NLO}^{\rm tot}$, the corresponding
experimental cross sections $\sigma_{\rm data}^{\rm tot}$, and the ratios
\begin{equation}
R=\frac{\sigma_{\rm NLO}^{\rm tot}}{\sigma_{\rm data}^{\rm tot}} 
\label{eq:R}
\end{equation}
of the theoretical predictions to the measured values. In the presented values for $R$, we
have added in quadrature the experimental and theoretical uncertainties. One can see from the
results shown in the table that within combined experimental and theoretical uncertainties, 
NLO pQCD with our factorization breaking scheme gives a good description of the integrated
cross section of diffractive dijet photoproduction in $ep$ scattering measured at HERA.
\begin{table}[h]
\caption{The theoretical and experimental values of the total integrated cross sections of
 diffractive dijet photoproduction in $ep$ scattering at HERA,  $\sigma_{\rm NLO}^{\rm tot}$
 and $\sigma_{\rm data}^{\rm tot}$, and their ratios $R$, see Eq.~(\ref{eq:R}).}
\begin{center}
\begin{tabular}{|c|c|c|}
\hline
H1, low-$E_T$ cut & H1, high-$E_T$ cut & ZEUS \\
\hline
$\sigma_{\rm data}^{\rm tot}=295 \pm 6(\rm stat.) \pm 58 (\rm syst.)$  pb & 
$\sigma_{\rm data}^{\rm tot} = 37 \pm 2(\rm stat.) \pm 8 (\rm syst.)$ pb  &
$\sigma_{\rm data}^{\rm tot} = 124^{+11}_{-5}$ pb
\\
\hline
$\sigma^{\rm NLO}_{\rm tot}=375^{+157}_{-81}$ pb & 
$\sigma^{\rm NLO}_{\rm tot}=51^{+15}_{-11}$ pb   &
$\sigma^{\rm NLO}_{\rm tot}=165^{+46}_{-34}$ pb
\\
$R=1.27^{+0.46}_{-0.29}$ & 
$R=1.38^{+0.37}_{-0.31}$ & 
$R=1.33^{+0.29}_{-0.21}$
\\
                   \hline
\end{tabular}
\end{center}
\label{table:sigma}
\end{table}%
Note that the central value of $R$ for the high-$E_T$ cut data is somewhat larger than that
for the low-$E_T$ cut, which can be explained by the fact that the unsuppressed charm quark
contribution becomes more prominent due to the QCD evolution of the photon PDFs.

\section{Conclusions}

We calculated the cross sections of diffractive dijet photoproduction in $ep$ scattering in
HERA kinematics using NLO perturbative QCD and a scenario of factorization breaking, which
assumes that only the resolved-photon contribution is suppressed. The suppression depended
on the parton flavor and the light-cone momentum fraction of partons in the photon. It was
absent for charm quarks, larger for gluons than for light quarks, and decreased with an
increase of the parton momentum. This model for factorization breaking in diffractive
QCD is based on the space-time picture of photon-hadron interactions and complies with the
good pQCD description of diffractive photoproduction of open charm in $ep$ scattering.
We compared our results with the available H1 and ZEUS data and found that various measured
distributions and the integrated cross sections can be reproduced by our calculations with
good accuracy. This agreement allows us to advocate our model as a viable alternative to the
purely phenomenological scheme based on a global suppression factor.

\begin{acknowledgements}

VG would like to acknowledge useful discussions of photon PDFs with W.~Vogelsang and
factorization breaking with M.~Strikman and to thank the Institut f\"ur Theoretische
Physik, Westf\"alische Wilhelms-Universit\"at M\"unster, where this work was performed,
for hospitality. The work of VG is partially supported by a grant of Deutscher Akademischer
Austauschdienst (DAAD). The work of MK is partially supported by the BMBF Verbundprojekt
05H2015 through grant 05H15PMCCA.

\end{acknowledgements}

\end{document}